\documentclass[final,5p,times,twocolumn,letterpaper]{elsarticle}

\usepackage{natbib}
\usepackage{amsthm}         
\usepackage{amsmath} 	   
\usepackage{amssymb}       
\usepackage{amsfonts}
\usepackage{graphicx} 	    
\usepackage{verbatim}
\usepackage{color}
\usepackage{float}

\begin{document}

\begin{frontmatter}

\title{NEC violation in mimetic cosmology revisited}

\author[add0]{Anna Ijjas\corref{cor1}}
\ead{aijjas@princeton.edu}
\author[add1]{Justin Ripley}
\author[add0,add1]{Paul J. Steinhardt}

\address[add0]{Princeton Center for Theoretical Science, Princeton 
University, Princeton, NJ 08544 USA}
\address[add1]{Department of Physics, Princeton University, Princeton, NJ 
08544, 
USA}

\cortext[cor1]{Corresponding author.}


\date{\today}

\begin{abstract}
In the context of Einstein gravity, if the null-energy condition (NEC) is satisfied, the energy density in expanding space-times always decreases while in contracting space-times the energy density grows and the universe eventually collapses into a singularity. In particular, no non-singular bounce is possible. It is, though, an open question if this energy condition can be violated in a controlled way, {\it i.e.}, without introducing pathologies, such as unstable negative-energy states or an imaginary speed of sound.
In this paper, we will re-examine the claim that the recently proposed mimetic scenario can violate the NEC without pathologies. We show that mimetic cosmology is prone to gradient instabilities even in cases when the NEC is satisfied (except for trivial examples). Most interestingly, the source of the instability is always the Einstein-Hilbert term in the action. The matter stress-energy component does not contribute spatial gradient terms but instead makes the problematic curvature modes dynamical.
We also show that mimetic cosmology can be understood as a singular limit of known, well-behaved theories involving higher-derivative kinetic terms and discuss ways of removing the instability.

\end{abstract}



\end{frontmatter}

\section{Introduction}

Traditionally, the null energy condition (NEC) is assumed in general relativity, high-energy physics, and cosmology. It implies that, for every null-vector $k^{\mu}$, the stress-energy tensor $T_{\mu\nu}$ obeys the inequality
\begin{equation}
\label{nec}
T_{\mu\nu} k^{\mu}k^{\nu}\geq 0\,.
\end{equation}
For perfect fluids, this criterion means that the sum of energy density $\rho$ and pressure $p$ remains non-negative. In a Friedmann-Roberston-Walker (FRW) space-time ($ds^2=-dt^2+a^2dx_idx^i$, where $a$ is the scale factor), if using reduced Planck units ($M_{\rm Pl}=1$), the Einstein equations simplify to
\begin{eqnarray}
\label{frw1}
3H^2 &= &\rho_{\text{tot}}, \\
\label{frw2}
 \dot{H} &=& - (\rho_{\text{tot}} + p_{\text{tot}})
 \,.
\end{eqnarray}
In an expanding universe, the total energy density or, equivalently, the Hubble parameter $H=\dot{a}/a>0$ (where dot denotes differentiation with respect to time $t$) always decreases if the NEC is satisfied. On the other hand, in a contracting universe, for which $H<0$, NEC satisfaction leads to continuous increase of the total energy density. In many cases, NEC violation is known to lead to pathologies, such as negative kinetic energy states (ghost) or imaginary speed of sound ($c_s^2<0$, gradient instability) \cite{Rubakov:2014jja}.
Recently, though, the question whether it is possible to violate the NEC without introducing these instabilities has received a fair amount of attention \cite{ArkaniHamed:2003uy,Creminelli:2006xe,Dubovsky:2005xd,Elder:2013gya,Koehn:2015vvy}. 

Settling the issue is particularly important for bouncing cosmologies, where it is assumed that the big bang is not a beginning but a bounce, connecting a cosmic phase of contraction to one of expansion \cite{Khoury:2001bz,Buchbinder:2007ad,Ijjas:2015zma}. The possibility of a contracting smoothing phase is intriguing because it only requires simple ingredients, such as a perfect fluid component or scalar field with equation of state $\epsilon = (3/2)(1+w)>3$ (where $w=p/\rho$), to smooth and flatten the cosmological background. In addition, (nearly) scale-invariant, super-horizon modes with small non-gaussianity can be generated during contraction that seed structure in the expanding, post-bounce universe and hence explain observations of the cosmic microwave background (CMB) \cite{McFadden:2005mq,Ijjas:2014fja}. Also, the physics of the contracting phase is well understood: on macroscopic scales, it is fully described by the classical Friedmann solutions of general relativity while on microscopic scales, it can be modeled using scalar fields and potentials. It is a great advantage of smoothing contraction that, unlike inflation, it does not lead to a multiverse or self-reproduction and does not involve any initial conditions problem. But the contracting phase has to end at some point and has to transit to the expanding phase of standard big-bang cosmology (bounce).  

To realize a cosmological bounce, two general strategies have been suggested: theories either involving a {\it singular} or a {\it non-singular} bounce. In the case of singular bounces, the energy density grows to reach the Planck scale. Consequently, understanding singular bounces requires some knowledge of or assumptions about quantum gravity \cite{Bars:2013vba,Gielen:2015uaa}.
In non-singular bouncing models, contraction stops at low energies and the transition to expansion occurs at a finite value of the scale factor, sufficiently far from the Planck scale, where the Hubble parameter is negative and finite. A successful transition to the high-energy expanding phase requires the Hubble parameter to grow, eventually hit zero, switch sign and continue to grow until a high enough energy $\sim H^2$ is reached for standard big-bang evolution to follow.  This approach has the advantage that it does not require knowledge of quantum gravity but one can safely rely on the low-energy effective theory. However, obviously, this type of non-singular bounce  involves a NEC violating form of stress-energy so that it stands or fails depending on whether it is possible to violate the NEC without pathologies. (Alternatively, one could, of course, try to introduce a modification to Einstein gravity that makes it possible to bounce without introducing a form of stress-energy that violates the NEC.) 

Recently, an interesting novel ansatz, {\it mimetic cosmology}, has been proposed that is supposed to violate the NEC and to avoid associated pathologies, such as gradient or ghost instabilities \cite{Chamseddine:2013kea,Chamseddine:2014vna}. In this paper, we revisit this claim and show that simple mimetic scenarios can indeed evade ghost instabilities but nevertheless have gradient instabilities even if the NEC is satisfied (except for trivial cases).  In Sec.~\ref{basics}, we start with briefly reviewing the basics of mimetic cosmology. Then, we derive the second-order action in spatially-flat and co-moving gauges in Sec.~\ref{stability}. 
We demonstrate in Sec.~\ref{limit} that the mimetic theory can be understood as a singular limit of known, well-behaved theories involving higher-order kinetic terms and briefly discuss ways of avoiding instabilities.
Finally, in Sec.~\ref{discussion}, we relate our result to earlier work that was done in Newtonian gauge and  stress the importance of analyzing stability using the action formalism with gauge-invariant quantities as opposed to considering only the perturbed equations of motion. 

\section{Basics of mimetic cosmology}
\label{basics}

In mimetic cosmology, the underlying idea is to perform a metric transformation,
\begin{equation}
g_{\mu\nu} = \left( \tilde{g}^{\alpha\beta}\partial_{\alpha}\phi\partial_{\beta}\phi \right) \tilde{g}_{\mu\nu}.
\label{MetricTransf}
\end{equation}
Here and throughout, $\phi$ is a scalar field, $g_{\mu\nu}$ is the physical metric, $\tilde{g}_{\mu\nu}$ is an auxiliary metric and we use reduced Planck units $M_{\rm Pl}^2= 1/(8\pi G_{\rm N})=1$ with $G_{\rm N}$ being Newton's constant. It is easy to see that the scalar field satisfies the mimetic constraint
\begin{equation}
\label{MimeticConstraint}
g^{\mu\nu}\partial_{\mu}\phi\partial_{\nu}\phi = -1.
\end{equation}
The transformation in Eq.~\eqref{MetricTransf} isolates the conformal degree of freedom in a covariant way and is a particular example of {\it singular} metric transformations \cite{Bekenstein:1992pj,Domenech:2015tca}.
 
In \cite{Golovnev:2013jxa} it has been shown that the mimetic transformation as defined in Eq.~\eqref{MetricTransf} is equivalent to introducing a Lagrange multiplier in the Einstein-Hilbert action for the physical metric $g_{\mu\nu}$, {\it i.e.},
\begin{equation}
\label{simple-mimetic}
S =  \int d^4x \sqrt{-g} \left( \frac{1}{2}R 
+ \lambda\left( g^{\mu\nu}\partial_\mu \phi\partial_\nu\phi+1 \right) \right)
,
\end{equation}
where $R$ is the Ricci scalar, $g$ the metric determinant and $\lambda$ the non-zero Lagrange multiplier. In both formulations, the mimetic constraint {\it parametrically} renders the scalar field $\phi$ to follow the background solution $\phi= t + \text{\it constant}$. Consequently, mimetic cosmology can either be considered as a modification of Einstein's gravity or as the introduction of a new form of stress-energy. In the following, we will embrace the latter point of view; in particular, because it makes extensions and generalizations of the mimetic idea possible. 

We shall consider the mimetic action as introduced in Ref.~\cite{Chamseddine:2014vna},
\begin{eqnarray}
\label{fullmimetic}
S = &\int& d^4x\sqrt{-g} \frac{1}{2} R \\
+&\int& d^4x\sqrt{-g} \left(\lambda\left( g^{\mu\nu}\partial_\mu \phi\partial_\nu\phi+1 \right) - V(\phi)\right)\nonumber\\
+ &\int& d^4x\sqrt{-g} \frac{\gamma}{2} (\Box \phi)^2 
,\nonumber
\end{eqnarray}
where $\gamma$ is a non-zero constant. In the context of mimetic gravity, the higher-order term $\propto \gamma$ has been introduced for purely phenomenological reasons, {\it i.e.}, to obtain dynamical spatial gradient terms in a simple way, and does not follow from the underlying principles of mimetic theory. Interestingly, though, the action in Eq.~\eqref{fullmimetic} coincides with the action for the IR-limit of projectable Ho\v{r}ava-Lifshitz gravity \cite{Ramazanov:2016xhp}. We will further comment on the connection between these two theories in the discussion section.

In a FRW space-time, the mimetic action in Eq.~\eqref{fullmimetic} admits the homogeneous background solution
\begin{eqnarray}
\label{mimetic-background}
3H^2 +  2 \dot{H} &=& \frac{2}{2-3\gamma}V(\phi) ,\\
\label{mimetic-background2}
\dot{H} &=& \frac{1}{1- 3\gamma }\lambda,
\\
\label{mimetic-background3}
\phi &=& t
.
\end{eqnarray}

Obviously, for appropriate values of the parameters $\lambda$ and $\gamma$, $\dot{H}$ can be positive such that the background solution violates the NEC. This is an interesting feature of mimetic gravity since it presents a novel way of NEC violation that would require relatively simple ingredients -- a singular metric transformation using a single scalar field.  NEC violation is often known, though, to involve pathologies, such as ghost or gradient instabilities. In the next section, we perform the linear stability analysis and show that the background corresponding to the action in Eq.~\eqref{fullmimetic} is not well-behaved under perturbations. 

\section{Linear stability analysis}
\label{stability}

 To identify the presence of any instabilities, we employ the action formalism and derive the linear theory for first-order, gauge-invariant quantities.
For this purpose, it is convenient to work with the ADM decomposition of the metric, 
\begin{equation}
\label{adm-decomp}
ds^2 = -N^2dt^2 + h_{ij}(N^idt + dx^i)(N^jdt + dx^j),
\end{equation}
where $N$ is the lapse, $N_i$ is the shift and $h_{ij}$ is the spatial metric that we use to raise and lower indices.
 Note that at zeroth order, $N=1$ and $N_i=0$ and $h_{ij}=a(t)\delta_{ij}$, the well-known FRW metric, and $\dot{\phi}=1$, the mimetic solution.
 
The ADM decomposition of the mimetic action as given in Eq.~\eqref{fullmimetic} is 
\begin{equation}
\label{ADMaction0}
S = \int d^4x\, \sqrt{h}\,N \,\mathcal{L}_{\rm ADM} 
\end{equation}
with the Lagrangian density 
\begin{eqnarray}
\label{ADMaction}
\mathcal{L}_{\rm ADM} &=& \frac{1}{2}\left(R^{(3)} + \frac{1}{N^2} \left(E_{ij}^2-E^2 \right)\right),\\
&+& \lambda \left( - \frac{1}{N^2}\dot{\phi}^2 + 2\frac{N^i}{N^2}\dot{\phi}\partial_i\phi+ g^{ij}\partial_i\phi\partial_j\phi + 1 \right)  
\nonumber\\
&+&  \frac{\gamma}{2}\frac{1}{h\,N^2}\Bigg(\partial_t\left(\sqrt{h} \frac{\,-\dot{\phi} + N^i\partial_i\phi}{N}\right) + \partial_i\left(\sqrt{h} \frac{N^i\dot{\phi}}{N} \right)+
\nonumber\\
&&\quad\, +\, \partial_i\left(\sqrt{h} Ng^{ij}\partial_j\phi \right)\Bigg)^2 - V(\phi)
\nonumber,
\end{eqnarray}
where $h$ is the spatial-metric determinant, $R^{(3)}$ is the three dimensional Ricci scalar, $g^{ij}=h^{ij}-N^iN^j/N^2$ is the inverse spatial metric, and $K_{ij} = -E_{ij}/N$ is the extrinsic curvature with
\begin{equation}
E_{ij}=\frac{1}{2}\left( \dot{h}_{ij}-\nabla_i N_j-\nabla_j N_i \right).
\label{extr.curv}
\end{equation}

To study the linear theory of scalar perturbations, we next have to choose a particular gauge to fix the remaining degrees of freedom.

\subsection{Spatially-flat gauge}
\label{sec-SF}

First, we shall derive the second-order action in spatially-flat gauge in which all spatial inhomogeneities are promoted to perturbations of the scalar field $\delta \phi \equiv \pi(t, {\bf x})$ while the spatial metric does not carry any perturbations, $h_{ij} = a^2\delta_{ij}$. Our gauge choice makes it straightforward to identify the instability and to relate our results to both Newtonian gauge in which the original calculation has been performed \cite{Chamseddine:2014vna} and to co-moving gauge in which the source of the instability can be most easily understood.

In spatially-flat gauge, expanding each term in the action in Eq.~\eqref{ADMaction} to second order in perturbations yields 
\begin{equation}
S^{(2)}_{\pi} = \int d^4x\, a^3 \mathcal{L}_{\pi}^{(2)}
\end{equation}
with
\begin{eqnarray}
\label{Lphi}
\mathcal{L}_{\pi}^{(2)} 
&=& \left(-\bar{\lambda} +\frac{9}{2}\gamma H^2 \right)\dot{\pi}^2   + \bar{\lambda}\frac{(\partial_k \pi)^2}{a^2} + \frac{3}{2}\dot{\bar{V}} H\pi^2
\\
&+& \left(-3H^2 - \bar{\lambda}+27\gamma H^2 \right) N_1^2 + \left(2\bar{\lambda} -27\gamma H^2 \right) \dot{\pi}N_1
\nonumber\\
&-& 2\left((1-3\gamma)HN_1 + \bar{\lambda}\pi + \frac{3}{2}\gamma\left( H\dot{\pi} -\dot{H}\pi \right)\right) \frac{\Delta\chi}{a^2}
\nonumber\\
&+& \frac{\gamma}{2}\left(\frac{\Delta\chi+\Delta\pi}{a^2}\right)^2+ 3\gamma H(2N_1-\dot{\pi}) \frac{\Delta\pi}{a^2}
\nonumber\\
&+&\frac{\gamma}{2}\,\Bigg(\;\ddot{\pi}^2
+\dot{N}_1^2 -2\ddot{\pi}\dot{N}_1 + 2\left(\dot{N}_1-\ddot{\pi} \right)\frac{\Delta\chi+\Delta\pi}{a^2}\Bigg)
\nonumber\\
&+& 3\gamma H\left( \ddot{\pi}\left(\dot{\pi} -3N_1\right) -2\dot{N}_1\left(\dot{\pi} -2N_1\right)\right)
\nonumber\\
&+& 2 \left(N_1-\dot{\pi}\right)\delta\lambda
\,,
\nonumber
\end{eqnarray}
where $\Delta\equiv\partial_i\partial^i$, $\delta \lambda = \lambda - \bar{\lambda}$ is the linear perturbation to the homogeneous Lagrange multiplier $\bar{\lambda}$, $N_1=N-1$ and $N_i=\partial_i\chi$, the linear perturbations to the lapse and shift, respectively, and we have used that on the mimetic background $\dot{V}=V,_{\phi} \dot{\phi}=V,_{\phi}$. (The bar over any quantity below refers to the unperturbed value.) Note, from Eq.~\eqref{Lphi}, we can immediately recover the result found in Ref.~\cite{Barvinsky:2013mea} that studied linear-order  perturbations of the mimetic scalar for $\gamma=0$ on flat space ($N_1=\Delta\chi=0$). In this case, the action in Eq.~\eqref{Lphi} reduces to its first line and the ghost instability is, indeed, avoidable for $\bar{\lambda}>0$, as suggested in~\cite{Barvinsky:2013mea}. However, what was missed is that avoiding the ghost instability comes at the cost of introducing a gradient instability!

At linear order, the equation for $\delta\lambda$ yields a new mimetic constraint,
\begin{equation}
\label{deltalambda}
N_1=\dot{\pi},
\end{equation}
such that time derivatives of the lapse $N_1$ do not carry any additional degree of freedom. 
Substituting the linear-order mimetic constraint into Eq.~\eqref{Lphi} and using the background solution in Eq.~\eqref{mimetic-background2} to eliminate $\bar{\lambda}$, the action simplifies to 
\begin{eqnarray}
\label{Lpi-simple}
\mathcal{L}_{\pi}^{(2)} 
&=& -\frac{3}{2} \left(2-3\gamma \right)H^2\dot{\pi}^2  \\
&+&\frac{1}{2}\left( \left(2-3\gamma \right)\dot{H}+3\gamma H^2\right)\frac{(\partial_k \pi)^2}{a^2} + \frac{3}{2}\dot{\bar{V}} H\pi^2
\nonumber
\\
&-& \left(\left(2- 3\gamma\right) \left( \dot{H}\pi+H\dot{\pi} \right) - \gamma\frac{\Delta\pi}{a^2}\right) \frac{\Delta\chi}{a^2} + \frac{\gamma}{2}\left(\frac{\Delta\chi}{a^2}\right)^2
\nonumber\\
&+& \frac{\gamma}{2}\left(\frac{\Delta\pi}{a^2}\right)^2 
.
\nonumber
\end{eqnarray}
The equation for $\chi$ follows immediately as
\begin{equation}
\label{chi-eq}
\gamma \frac{\Delta\chi}{a^2} = \left(2-3\gamma\right)\left( H\dot{\pi}+\dot{H}\pi \right)  - \gamma \frac{\Delta\pi}{a^2}.
\end{equation}
From the $\chi$-equation, it is straightforward to recover the results found by Chamseddine et al. in Ref.~\cite{Chamseddine:2014vna} by way of canonical coordinate transformation,  as we show in the Appendix~\ref{appendix}.

If $\gamma = 0$ and we have no higher-derivative term in the mimetic action~\eqref{fullmimetic}, the $\chi$-equation renders
\begin{equation}
\label{chi-eq-gamma0}
 H\dot{\pi}+\dot{H}\pi =0,
\end{equation}
and the second-order action reduces to 
\begin{eqnarray}
\label{lambda=zero}
S^{(2)}_{\pi, \gamma=0} &=& \int d^4x\, a^3  \left(-3\left(\partial_t(H\pi)\right)^2+\left(H\frac{\partial_k\pi}{a}\right)^2\right)
.
\end{eqnarray}
Here, we have used Eq.~\eqref{chi-eq-gamma0}, the background solution in Eq.~(\ref{mimetic-background}-\ref{mimetic-background2}) and, integrated by parts to find $a\bar{\lambda}(\partial_k\pi)^2=aH^2(\partial_k\pi)^2$ and $(3/2)\dot{V}\pi^2=3\dot{H}^2\pi^2+6H\dot{H}\pi\dot{\pi}$. 
It is immediately apparent that, independent of the sign of $\bar{\lambda}$, both the kinetic and gradient terms for the gauge-invariant quantity $H\pi$ carry wrong sign, indicating the presence of instabilities even if the background satisfies the NEC, {\it i.e.}, $\dot{H}=\bar{\lambda}\geq0$. In our convention, ghost instability corresponds to a negative coefficient of the kinetic term $\dot{\pi}^2$ in the action and gradient instability corresponds to a positive coefficient of the gradient term $(\partial_k\pi)^2$. 
(Note that $\pi$ is not gauge-invariant!) But, from Eq.~\eqref{chi-eq-gamma0}, it follows the combination $H\pi$ has no time-dependence; hence, although the instabilities are present, they cannot grow. 

If $\gamma \neq 0$, substituting the equation for $\chi$ into Eq.~\eqref{Lpi-simple} and integrating by parts, second-order action in spatially-flat gauge takes the simple form 
\begin{equation}
\label{Lpi-gammaneq0}
S^{(2)}_{\pi} = \int d^4x\, a^3 \left(-\frac{2-3\gamma}{\gamma}\left(\partial_t(H\pi)\right)^2+\left(H\frac{\partial_k\pi}{a}\right)^2\right)
\,.
\end{equation}
Hence, in Eq.~\eqref{Lpi-gammaneq0}, we see that the theory has a gradient instability for all backgrounds and it may or may not have a ghost instability, depending on the sign of $(2-3\gamma)/\gamma$. Note that, after integrating out $\chi$, the final action for the $\gamma=0$ case as given in Eq.~\eqref{lambda=zero} cannot be obtained as the $\gamma \to 0$ limit of Eq.~\eqref{Lpi-gammaneq0}. That means, the instability appears for any non-zero $\gamma$.  

\subsection{Co-moving gauge}
\label{sec-comoving}

To better understand the source of the gradient instability, it proves useful to repeat parts of the linear stability analysis in co-moving gauge and then to compare our results in both gauges. 

In co-moving gauge, all spatial inhomogeneities are promoted to the metric, 
\begin{equation}
h_{ij} = a^2(t)e^{2\zeta(t,{\bf x})}\delta_{ij},
\label{hzeta}
\end{equation}
where $\zeta$ is the gauge-invariant, co-moving curvature perturbation; while the scalar field remains homogeneous and does not carry any perturbations,
\begin{equation}
\pi_{\zeta}=0.
\label{pizeta}
\end{equation}

In co-moving gauge, the mimetic action defined in Eq.~\eqref{ADMaction} takes the form 
\begin{equation}
\label{ADM-zeta}
S_{\zeta} = \int d^4x\, a^3e^{3\zeta}N \mathcal{L}_{\zeta} \end{equation}
with
\begin{eqnarray}
\label{ADM-zeta-lagrangian}
 \mathcal{L}_{\zeta} &=& - e^{-2\zeta}\left(2\frac{\Delta \zeta}{a^2} + \frac{(\partial_k \zeta)^2}{a^2} \right) -3\frac{1}{N^2}(H+\dot{\zeta})^2
\\
 &+& 2\frac{e^{-2\zeta}}{N^2}\left( (H+\dot{\zeta})\frac{\Delta\chi}{a^2} + H\frac{\partial^k\zeta\partial_k\chi}{a^2}\right)
\nonumber
\\
&+& \lambda\left( -\frac{\dot{\bar{\phi}}^2}{N^2}+1 \right) - V(\bar{\phi})
\nonumber\\
\nonumber
&+& \frac{\gamma}{2}\Bigg(-3\frac{\dot{\bar{\phi}}}{N^2}\left(H+\dot{\zeta}\right) - \frac{\ddot{\bar{\phi}}}{N^2} -\frac{\dot{\bar{\phi}}}{N}\partial_t\left(\frac{1}{N}\right) +
\\
&&\quad + e^{-2\zeta}\frac{\dot{\bar{\phi}}}{N}\left(\frac{1}{N}\frac{\Delta\chi}{a^2} + \frac{\partial^k\zeta\partial_k\chi }{a^2}+\partial^k\left(\frac{1}{N}\right)\frac{\partial_k\chi }{a^2}\right)
\Bigg)^2
; \nonumber
\end{eqnarray}
here, the first two lines are the contribution of the gravitational sector.

We shall expand this action around the mimetic background given through Eq.~\eqref{mimetic-background3} to second order in perturbations. 
As in Sec.~\ref{sec-SF}, at second-order in perturbations, the $\lambda$-sector of the Lagrangian, 
\begin{equation}
\label{mimeticConstraint-zeta0}
\left( - \bar{\lambda}N_1+ 2\delta\lambda\right)N_1,
\end{equation}
yields a first-order mimetic constraint resulting from the linearized field equation for $\delta \lambda$, 
\begin{equation}
\label{N1zeta}
N_1=0.
\end{equation}
Note that Eq.~\eqref{N1zeta} is consistent with the expression for $N_1$ in spatially-flat gauge, Eq.~\eqref{deltalambda}: Under an infinitesimal coordinate change $\xi_0$, the lapse transforms as $N_1|_{g_1}= N_1|_{g_2}-\dot{\xi}^0$, where ${}|_{\rm g}$ denotes that a quantity is evaluated in gauge $g$ \cite{Bardeen:1980kt}.
For transformations between the spatially-flat and co-moving gauges on the mimetic background ($\dot{\bar{\phi}}=1$), the defining coordinate change $\xi^0_{\pi\to\zeta} =-\pi$ can be identified, for example, by using the fact that scalar-field perturbations transform as $\pi=\pi_{\zeta}-\xi^0_{\pi\to\zeta}\dot{\bar{\phi}}$ and $\pi_{\zeta}=0$ by definition. Hence, $N_1|_{\pi}= N_1|_{\zeta}+\dot{\pi}=\dot{\pi}$ as found in Eq.~\eqref{deltalambda}.

Substituting Eq.~\eqref{N1zeta} into Eq.~(\ref{ADM-zeta-lagrangian}), the expressions significantly simplify and the second-order action in co-moving gauge takes the form
\begin{equation}
S^{(2)}_{\zeta} = \int d^4x\, a^3 \mathcal{L}_{\zeta}^{(2)}
\end{equation}
with
\begin{eqnarray}
\label{L2zeta}
\mathcal{L}_{\zeta}^{(2)} 
&=& -\left(3 - \frac{9}{2}\gamma\right)\dot{\zeta}^2 +\frac{(\partial_k\zeta)^2}{a^2} 
\\
\nonumber
&+& \left(2-3\gamma\right)\dot{\zeta}\frac{\Delta\chi}{a^2} + \frac{\gamma}{2} \left(\frac{\Delta\chi}{a^2}\right)^2  
\\
&+& \frac{9}{2} \left( 3\frac{3\gamma-2}{2}H^2 -\bar{V}\right)\zeta^2   + 9\frac{3\gamma-2}{2}H\, \partial_t\zeta^2  
\,.
\nonumber
\end{eqnarray}
The $\chi$-equation reduces to the simple relation
\begin{equation}
\label{chi-zeta}
\gamma\frac{\Delta\chi}{a^2} = -(2-3\gamma) \dot{\zeta}.
\end{equation}
Note that, if $\gamma=0$, the equation for $\chi$ renders $\zeta$ non-dynamical (that is, $\dot{\zeta}=0$ precisely). 

Integrating out $\chi$ and using the background solution in Eq.~\eqref{mimetic-background} to eliminate $\bar{V}$ yields the second-order action for $\zeta$:
\begin{equation}
\mathcal{S}^{(2)}_\zeta=\int d^4x a^3\left( -A_{\gamma}\dot{\zeta}^2+\frac{(\partial_k \zeta)^2}{a^2} \right),
\label{L2zeta}
\end{equation}
with the kinetic coefficient 
\begin{equation}
A_{\gamma} =  \begin{cases}
  3    & \text{if } \gamma=0\,, \\
  \frac{2-3\gamma}{\gamma}    & \text{if } \gamma \neq 0\,.\;
\end{cases}
\label{L2zeta-coeff}
\end{equation}
in agreement with our results in spatially-flat gauge as given in Eqs.~(\ref{lambda=zero}-\ref{Lpi-gammaneq0}); the consistency of both calculations can be easily verified by applying the well-known transformation rule for scalar perturbations to the spatial curvature $\psi$.   By definition, $\psi|_{\pi}=0$, $\psi|_{\zeta}=-\zeta$, and $\psi$ transforms as $\psi|_{g1}=\psi|_{g2}+H\xi^0_{g1\to g2}$. With $\xi^0_{\pi\to\zeta} = -\pi$ as shown above, $\zeta=-H\pi$. 

(Note that the spatial metric $h_{ij}$ introduced in Eq.~\eqref{adm-decomp} can be related to $\psi$, for example, via $h_{ij}=a^2(t)\left((1-2\psi|_{\zeta})\delta_{ij}+ \alpha_{ij} + \mathcal{O}(2)+...\right)$, where $\alpha_{ij}$ is a traceless and transverse tensor and does not carry  scalar degrees of freedom so we can neglect it. Note, though, that, in contrast to $\psi$, $h_{ij}$  is gauge independent.)

For all values of $\gamma$ (including $\gamma=0$), the first-order mimetic constraint leads to notable consequences: 
\begin{itemize}
\item[-] in co-moving gauge, the $\lambda$-sector does not contribute to the second-order action; 
\item[-] the remaining scalar sector only contributes to the kinetic term $\sim \dot{\zeta}^2$ but it does not contribute to the gradient term of $\zeta$; instead, all contribution to $(\partial_k\zeta)^2$ comes from the gravitational Einstein-Hilbert term in the action which is the source of the gradient instability. Without the higher-order term ($\gamma=0$), the theory formally suffers from both ghost and gradient instabilities, both coming from the Einstein-Hilbert term in the action, but the co-moving curvature remains non-dynamical so that the instability cannot grow. 
Adding the higher-order kinetic term ($\gamma\neq0$) can eliminate the ghost for appropriate choice of $\gamma$ and make $\zeta$ become dynamical but it cannot alleviate the gradient instability.  
\end{itemize}
It is worth emphasizing that these features could easily be recognized by employing the action formalism and performing the perturbation calculation in co-moving gauge, using gauge-invariant quantities. Working in spatially-flat gauge leads to the same conclusion, as it should, but the ultimate source of the instability is obscured. Working at the level of the perturbed equations of motion, on the other hand, can lead to wrong conclusions about the instability.

\section{Mimetic cosmology as a singular limit of higher-derivative theories}
\label{limit}

In the previous section, we have shown that gradient instabilities arise within mimetic cosmology if we add higher-order kinetic terms. Without higher-order kinetic terms, on the other hand, curvature modes do not get excited. To better understand the nature of both mimetic theory and the associated gradient instability, we next demonstrate that the mimetic action in Eq.~\eqref{fullmimetic} can be recovered as the limit of known higher-derivative theories. 

We shall consider the following action, 
\begin{eqnarray}
\label{hD10}
S &= &\int d^4x\sqrt{-g} \frac{1}{2} R \\
&+ &\int d^4x\sqrt{-g} \left(\lambda \left( \partial_{\mu}\phi\partial^{\mu}\phi + 1 \right) - \frac{\xi}{2}\lambda^2   - V(\phi)\right)
\nonumber\\
&+ &\int d^4x\sqrt{-g} \frac{\gamma}{2} (\Box \phi)^2 
,\nonumber
\end{eqnarray}  
where $\xi$ and $\gamma$ are constants and $\lambda$ is a dynamical variable described by the equation of motion
\begin{equation}
\label{lambda-intout}
\xi \lambda= \left(\partial_{\mu}\phi\partial^{\mu}\phi + 1\right).
\end{equation}

If $\xi=0$, $\lambda$ acts as a Lagrange multiplier, the $\lambda$-equation becomes the mimetic constraint and the action reduces to the mimetic action in Eq.~\eqref{fullmimetic}.

If $\xi\neq0$, substituting the expression for $\lambda$ in Eq.~\eqref{lambda-intout} into the original action in Eq.~\eqref{hD1}, we find 
\begin{eqnarray}
\label{hD1}
S &= &\int d^4x\sqrt{-g} \frac{1}{2} R \\
&+ &\int d^4x\sqrt{-g} \left(\frac{1}{2\xi} \left( \partial_{\mu}\phi\partial^{\mu}\phi + 1 \right)^2   - V(\phi)\right)
\nonumber\\
&+ &\int d^4x\sqrt{-g} \frac{\gamma}{2} (\Box \phi)^2 
\,.
\nonumber
\end{eqnarray}
This action represents a particular example of well-known $P(X)$-theories (where $X\equiv(\partial\phi)^2/2$) with additional higher-derivative terms, such as $k$-inflation \cite{ArmendarizPicon:1999rj}, $k$-essence \cite{ArmendarizPicon:2000ah}, or ghost condensate \cite{ArkaniHamed:2003uy}. If $ \gamma=0$, $P(X)$-theories encounter gradient instabilities on NEC violating backgrounds but are well-behaved if the NEC is satisfied. If $ \gamma\neq0$, $P(X)$-theories suffer from Ostrogradski instability that cannot be removed without altering the theory \cite{Chen:2012au}.  

First, we show that the mimetic background can be recovered as a {\it continuous} limit of Eq.~\eqref{hD1} as $\xi\to0$.

\subsection{Mimetic background as a smooth limit for $\xi\to0$}

The action in Eq.~\eqref{hD1} admits the FRW background solution
\begin{eqnarray}
\label{PX-bgr2}	
\frac{2-3\gamma\dot{\phi}^2}{2}\left(3H^2+2\dot{H}\right) &=& - \frac{1}{2\xi}\left(\dot{\phi}^2-1\right)^2 + V(\phi)
\\
&+& \gamma\left( 6H\ddot{\phi}\dot{\phi}  + \dot{\phi}\dddot{\phi}+\frac{1}{2}\ddot{\phi}^2\right)\,,
\nonumber
\end{eqnarray}
\begin{eqnarray}
\label{HdotPX}
\left(1-3\gamma\dot{\phi}^2 \right) \dot{H} &=& - \frac{\dot{\phi}^2}{\xi}\left( \dot{\phi}^2 -1\right)+ \gamma \dot{\phi}\left( \dddot{\phi}+3H\ddot{\phi}\right)
 \,.
\end{eqnarray}
In the limit of $\xi\to0$, the scalar field must satisfy the mimetic constraint in Eq.~\eqref{MimeticConstraint},
\begin{equation}
\label{xito0phi}
\dot{\phi}^2-1=0,
\end{equation}
as otherwise the total energy density ($\propto H^2$) would diverge. With the corresponding (mimetic) solution $\phi=t+{\rm constant}$, Eq.~\eqref{PX-bgr2} reduces to its mimetic counterpart Eq.~\eqref{mimetic-background},
\begin{equation}
\frac{2-3\gamma}{2}\left(3H^2+2\dot{H}\right) =   V(\phi)\,.
\end{equation} 
Finally, it is straightforward to recover Eq.~\eqref{mimetic-background2} from Eq.~\eqref{HdotPX}. Using Eq.~\eqref{lambda-intout},
\begin{equation}
\label{lambda-intoutbr}
\dot{\phi}^2=1-\xi \lambda
\,,
\end{equation}
and keeping terms to leading order in $\xi$, the $\dot{H}$-equation simplifies to its mimetic counterpart 
\begin{equation}
\left(1-3\gamma \right) \dot{H}= \lambda\,;
\end{equation}
Hence, as claimed, the mimetic background can be recovered as a continuous limit of the higher-derivative theory given through Eq.~\eqref{hD1}.

Notably, this is not the case for the perturbed action. We will next show that the second-order mimetic action is a {\it singular limit} of Eq.~\eqref{hD1} as $\xi\to0$. By singular limit, we mean the stability behavior for $\xi=0$ undergoes a discontinuous jump compared to finite $\xi$.

\subsection{Second-order mimetic action as a singular limit for $\xi\to0$}

Similar to Sec.~\ref{sec-comoving}, we perform the linear stability analysis by expanding the action in Eq.~\eqref{hD1} around the homogeneous background solutions derived from Eqs.~(\ref{PX-bgr2}-\ref{HdotPX}) to quadratic order in perturbations, using ADM variables. In co-moving gauge, the second-order action  is	
\begin{equation}
\label{limActExpand1}
	S_{\zeta}^{(2)} = \int d^4x\, a^3 \mathcal{L}_{\zeta}^{(2)}\,,
\end{equation}
where the Lagrangian density is
\begin{eqnarray}
\label{limLagExpand1}
\mathcal{L}_{\zeta}^{(2)}&=& -3\left( 1-\frac{3}{2}\gamma\dot{\phi}^2  \right) \dot{\zeta}^2+ \frac{(\partial_k\zeta)^2}{a^2}
\\
 &+& 2\left( -H N_1 + \gamma\left(\ddot{\phi}+3H\dot{\phi}\right)\dot{\phi} N_1 + \frac{\gamma}{2}\dot{\phi}^2\dot{N}_1 \right)\frac{\Delta\chi}{a^2}
 \nonumber\\
&+&\left( 2-3\gamma\dot{\phi}^2\right)\dot{\zeta}\frac{\Delta\chi}{a^2}
+\frac{\gamma}{2}\dot{\phi}^2\left(\frac{\Delta\chi}{a^2} \right)^2  + \frac{\gamma}{2}\dot{\phi}^2\dot{N}_1^2   
\nonumber \\
&+& \left(- 2 \frac{\Delta\zeta}{a^2} + 3H\left(2-3\gamma\dot{\phi}^2 \right)\dot{\zeta} + 3 \gamma \dot{\phi}^2 \ddot{\zeta}\right)N_1
\nonumber \\
&+& \Bigg(-3H^2  + \frac{1}{\xi}\left(3\dot{\phi}^4-\dot{\phi}^2\right)+\nonumber\\ 
&& \quad+ \,\gamma \left( \ddot{\phi}^2+9H^2\dot{\phi}^2- 2\dddot{\phi}\dot{\phi} -6\dot{H}\dot{\phi}^2 \right) \Bigg)\, N_1^2  
\nonumber;
\end{eqnarray}
here we used the background equations to eliminate terms $\propto~\zeta N_1$ and $\propto~\zeta^2$. 
Varying the action with respect to $\Delta \chi$ leads to the constraint equation
\begin{eqnarray}
\label{momConst1}
\gamma \dot{\phi}^2\frac{\Delta\chi}{a^2} & = &-\left( 2-3\gamma\dot{\phi}^2\right)\dot{\zeta}\\
&+& 2\left( H-\gamma\dot{\phi}\left(\ddot{\phi}+3H\dot{\phi}\right) \right)N_1 - \gamma\dot{\phi}^2\dot{N}_1
.\nonumber
\end{eqnarray}

If $\gamma=0$, the $\chi$-equation yields a closed expression for the lapse perturbation,
\begin{equation}
\label{chi0-xi}
N_1=\frac{\dot{\zeta}}{H}\,,
\end{equation}
and the second-order action 
\begin{eqnarray}
\label{limLagExpand1-00}
S_{\zeta, \gamma=0}^{(2)} &=& \int d^4x a^3 \Bigg(-3\dot{\zeta}^2
+ \frac{(\partial_k\zeta)^2}{a^2} - 2 \frac{\Delta\zeta}{a^2}N_1
\\
&&\qquad
+\,6HN_1\dot{\zeta} +\frac{1}{\xi}\dot{\phi}^2\left(3\dot{\phi}^2-1\right) N_1^2 
-3H^2N_1^2 \Bigg)
\nonumber
\end{eqnarray}
reduces to
\begin{eqnarray}
\label{limLagExpand1-0}
S_{\zeta, \gamma=0}^{(2)} = \int d^4x a^3
\left( \frac{1}{\xi} \dot{\phi}^2\frac{3\dot{\phi}^2-1}{H^2} \dot{\zeta}^2 +\frac{\dot{H}}{H^2} \frac{(\partial_k\zeta)^2}{a^2} \right)
\,.
\end{eqnarray}
Obviously, this expression differs from the second-order mimetic action $S^{(2)}_{\zeta, \gamma=0}$ in Eq.~\eqref{L2zeta}, corresponding to $\xi=0$. This is an example of what we mean by a singular limit of the action in Eq.~\eqref{limLagExpand1-0}. In this case, the root cause is that Eq.~\eqref{hD1} alone is not equivalent to Eq.~\eqref{hD10}. It is only equivalent if one additionally imposes the constraint in Eq.~\eqref{lambda-intout}.
With this additional constraint, we recover the perturbed mimetic action by deriving the expression for $\delta\lambda$ from Eq.~\eqref{lambda-intout},
\begin{equation}
\label{lambda-intout-2ndO}
\xi \delta\lambda= \dot{\phi}^2N_1=\left(1-\xi\bar{\lambda}  \right)N_1\,.
\end{equation}
Evaluating the $\delta\lambda$-equation for $\xi=0$ leads to the mimetic constraint on the lapse: $N_1=0$; the action in Eq.~\eqref{limLagExpand1-00} reduces to its mimetic counterpart in Eq.~\eqref{L2zeta} and the $\chi$-equation~\eqref{momConst1} renders $\dot{\zeta}=0$. 
On the other hand, for finite $\xi$, $N_1=\dot{\zeta}/H \neq 0$ as given in Eq.~\eqref{chi0-xi} so terms proportional to the lapse $N_1$ and its square $N_1^2$ in Eq.~\eqref{limLagExpand1-00} can counteract the wrong-sign terms $\propto\dot{\zeta}^2, \propto (\partial_k\zeta)^2$ resulting from the Einstein-Hilbert part of the original action in Eq.~\eqref{hD1}.
This is how it is possible that, for $\xi\neq0$, the perturbations are stable on NEC satisfying backgrounds with $3\dot{\phi}^2>1$ (no ghost); otherwise, the linear-order perturbations are unstable (ghost and/or gradient instability) -- the well-known feature of $P(X)$ theories. 

If $\gamma\neq0$, substituting the expression for $\Delta\chi/a^2$ from Eq.~\eqref{momConst1} into the action in Eq.~\eqref{limLagExpand1} and integrating by parts yields
\begin{eqnarray}
\label{limActExpand2}
\mathcal{L}_{\zeta}^{(2)}&=& -\frac{2-3\gamma\dot{\phi}^2}{\gamma\dot{\phi}^2}  \dot{\zeta}^2+ \frac{(\partial_k\zeta)^2}{a^2}
\\
&+& 2\left( \ddot{\zeta}-  \frac{\Delta\zeta}{a^2} + \frac{1}{\gamma\dot{\phi}^2}\left(H\left(2-3\gamma\dot{\phi}^2\right) - 2\gamma\ddot{\phi}\dot{\phi}\right) \dot{\zeta} \right)N_1
\nonumber \\
&-& \Bigg(2\left(\frac{1}{\gamma\dot{\phi}^2}-3\right)H^2  +\left(1+3\gamma\dot{\phi}^2\right) \dot{H} - \frac{1}{\xi}\left(3\dot{\phi}^4-\dot{\phi}^2\right)
\nonumber\\ 
&& \quad \, - 4H\frac{\ddot{\phi}}{\dot{\phi}} + \gamma \dot{\phi}\left( 3H\ddot{\phi}+ \dddot{\phi}  \right) \Bigg)\, N_1^2  
\nonumber;
\end{eqnarray}
Varying with respect to the lapse $N_1$, we write the Hamiltonian constraint as
\begin{equation}
\label{HamConst1}
\bar{B}(t)\, N_1=\ddot{\zeta} -  \frac{\Delta\zeta}{a^2} + \frac{\left(2-3\gamma\dot{\phi}^2 \right)H - 2\gamma\ddot{\phi}\dot{\phi}}{\gamma\dot{\phi}^2} \dot{\zeta} \,,
\end{equation}
with the time dependent coefficient  
\begin{eqnarray}
\label{HamConst11}
\bar{B}(t) &=& - \frac{1}{\xi}\dot{\phi}^2\left(3\dot{\phi}^2-1\right) 
+ 2\,\frac{1-3\gamma\dot{\phi}^2}{\gamma\dot{\phi}^2}H^2
\\   
&+& \left(1+3\gamma\dot{\phi}^2\right) \dot{H} 
- 4H\frac{\ddot{\phi}}{\dot{\phi}} + \gamma \dot{\phi}\left( \dddot{\phi}+3H\ddot{\phi}  \right)\,. 
\nonumber
\end{eqnarray}
Finally, substituting the expression for $N_1$, the second-order action for $\zeta$ defined through Eq.~\eqref{limActExpand1} is given by the Lagrangian density
\begin{eqnarray}
\label{actionZeta}
\mathcal{L}_{\zeta}^{(2)} &= &  -\frac{2-3\gamma\dot{\phi}^2}{\gamma\dot{\phi}^2}  \dot{\zeta}^2+ \frac{(\partial_k\zeta)^2}{a^2} 
\\
&+ & \frac{1}{\bar{B}(t)} 
\left(\ddot{\zeta} -  \frac{\Delta\zeta}{a^2} + \frac{\left(2-3\gamma\dot{\phi}^2 \right)H - 2\gamma\ddot{\phi}\dot{\phi}}{\gamma\dot{\phi}^2} \dot{\zeta}\right)^2
\,.
\nonumber 
\end{eqnarray} 
The higher-order time derivative $\ddot{\zeta}^2$ and the lack of any additional constraints on $\zeta$ indicates an Ostrogradski instability \cite{Chen:2012au}, as mentioned above. It is easy to see that for $\xi\to0$, $1/\bar{B}(t)\to0$, or alternatively, $N_1\to0$, and the action reduces to its mimetic counterpart in Eq.~\eqref{L2zeta} for $\gamma\neq0$. In this sense, the mimetic action is a continuous limit of the action in Eq.~\eqref{actionZeta}.  On the other hand, the transition from the Ostrogradski instability to the (mimetic) gradient instability is a discontinuous jump. For any $\xi\neq0$, the higher-order derivative term and, hence, the Ostrogradski instability is present but it vanishes when $\xi=0$, {\it i.e.}, $\xi=0$ is a singular limit.

\section{Discussion}
\label{discussion}

In this paper, we have performed a complete stability analysis of the recently introduced mimetic theory. We were particularly interested in the question whether mimetic cosmology opens up a new way to stably violate the null energy condition.  

We have found that the mimetic theory suffers from a gradient instability even if the null energy condition is satisfied. We have shown that the source of the gradient instability is the gravitational Einstein-Hilbert term in the action while the linear-order mimetic constraint prevents the scalar-field sector from contributing additional gradient terms that might stabilize the theory. 

In addition, we have demonstrated that the second-order mimetic action is a {\it singular} limit of known higher-derivative theories.  In particular, we have shown that even though for the higher-derivative theory the background equations of motion smoothly converge to the mimetic equations of motion in the limit $\xi\to0$, the linear stability behavior undergoes a discontinuous jump in the same limit.  This finding reveals that the mimetic theory is truly distinct from these higher-derivative theories.
This lesson is important in considering other theories that approach the mimetic in some limit.  For example, it has been noted that the projectable Ho\v{r}ava-Lifshitz gravity  approaches our mimetic action in the IR-limit \cite{Ramazanov:2016xhp}.  What we have shown here indicates that one should do an independent stability analysis  in the regime where gradient instabilities become dangerous, taking careful account of any additional contributions that are different from the mimetic theory.  See for example \cite{Koyama:2009hc,Cerioni:2010uz}. 

To perform our stability analysis, we focused on a particular mimetic action introduced in Ref.~\cite{Chamseddine:2014vna} but our results can easily be generalized.  For example, our interpretation of the mimetic theory as a singular limit of known higher-derivative theories opens up possible directions for future work, such as modifying the Lagrange-multiplier constraint in a way that eliminates the instability during NEC violation.

{\bf Acknowledgements.} We thank Slava Mukhanov for suggesting that we consider NEC violation in mimetic cosmology and Lasha Berezhiani for valuable contributions during the initial stages of the project. This research was partially supported by the U.S. Department of Energy under grant number DE-FG02-91ER40671.

\appendix
\section{Recovering the results in Ref.~\cite{Chamseddine:2014vna} from Eq.~\eqref{chi-eq}}
\label{appendix}

We show here that it is straightforward to recover the results found by Chamseddine et al. in Ref.~\cite{Chamseddine:2014vna}, where the linear-order mimetic perturbations were analyzed  in Newtonian gauge, at the level of the equations of motion.
In Newtonian gauge, the perturbed metric takes the form
\begin{equation}
ds^2 = - (1+2\Psi_N)dt^2 + (1-2\Phi_N)\,a^2(t)\delta_{ij}dx^idx^j\,.
\end{equation} 
Here, $\Psi_N$ is the Newtonian gravitational potential and $\Phi_N$ is the Newtonian spatial curvature. Notably, in the absence of anisotropic stress, $\Psi_N = \Phi_N$.
For an infinitesimal coordinate change $\xi^0$, the canonical transformation rules between spatially-flat and Newtonian gauges read as follows \cite{Bardeen:1980kt},
\begin{eqnarray}
\Psi|_{\pi} &=& \Psi_N + H \xi^0_{\pi\to N}\,,
\\ 
\chi|_{\pi} &=& \chi|_N + \xi^0_{\pi\to N}\,,
\\
\pi &=& \pi|_N-\xi^0_{\pi\to N}\dot{\bar{\phi}}
\,.
\end{eqnarray}
By definition, $\Psi|_{\pi} =0$ and $\chi|_N=0$, and the mimetic constraint in Eq.~\eqref{MimeticConstraint} renders $\dot{\bar{\phi}}=1$ and $\Psi_N= \dot{\pi}|_N$.
It follows, 
\begin{eqnarray} 
\chi|_{\pi} &=&-\frac{\dot{\pi}|_N}{H}\,,
\\
\pi &=& \pi|_N + \frac{\dot{\pi}|_N}{H}\,.
\end{eqnarray}
Substituting into Eq.~\eqref{chi-eq}, we find the equation for the Newtonian scalar-field perturbation $\pi|_N$,
\begin{equation}
\label{chi-eq-Newt}
 \left(2-3\gamma\right)\left( \ddot{\pi}|_N + H\dot{\pi}|_N +\dot{H}\pi|_N \right)  - \gamma \frac{\Delta\pi|_N}{a^2}=0 
 \,;
\end{equation}
in agreement with the result found in Ref.~\cite{Chamseddine:2014vna}.
Note, though, that the equation for $\pi_N$ is {\it not} determinative with respect to the stability of the physical metric perturbations, since the equation of motion only indicates the {\it relative} sign of kinetic and gradient terms of $\pi|_N$ and is insensitive to their absolute sign. Moreover,  the Newtonian scalar-field perturbation $\pi|_N$ is not gauge-invariant. One of the important lessons of the present paper is the advantages of employing canonical, gauge-invariant quantities and reading off the overall sign of the perturbed action in order to determine the existence and nature of instabilities.

\bibliographystyle{apsrev}
\bibliography{mimeticJ}

\end{document}